\documentclass{iopart}
\usepackage{graphicx}

\def\be{\begin{equation}}
\def\ee{\end{equation}}
\def\bea{\begin{eqnarray}}
\def\eea{\end{eqnarray}}

\begin{document}

\title{Making $h(t)$ for LIGO}

\author{Xavier Siemens$^1$, Bruce Allen$^1$, Jolien Creighton$^1$, 
Martin Hewitson$^2$ and Michael Landry$^3$}
\address{$^1$ Center for Gravitation and Cosmology, Department of Physics, 
University of Wisconsin - Milwaukee, P.O.Box 413, Wisconsin 53201, USA.}
\address{$^2$\ Max-Planck-Institut f\"ur Gravitationsphysik
  (Albert-Einstein-Institut) und Universit\"{a}t Hannover,
  Au\ss enstelle Hannover,
  Callinstr. 38, 30167~Hannover, Germany.}
\address{$^3$ LIGO Hanford Observatory, PO Box 159, Richland, WA 99352, USA.}

\ead{siemens@gravity.phys.uwm.edu}

\begin{abstract}
The conversion of the read-out from the anti-symmetric port of the
LIGO interferometers into gravitational strain has thus far been
performed in the frequency domain. Here we describe a conversion in
the time domain which is based on the method developed by GEO. We
illustrate the method using the Hanford 4km interferometer during the
second LIGO science run (S2).
\end{abstract}

\pacs{04.80.Nn,95.55.Ym, 95.75.Wx}

\maketitle{}

\section{Introduction}

The LIGO interferometers \cite{abr,barish,det_desc} are part of a
world-wide network of gravitational wave detectors constructed to
detect waves from astrophysical sources such as spinning neutron
stars, coalescing binary black holes and neutron stars, and
supernovae.

Each of the two arms of the LIGO interferometers forms a resonant
Fabry-Perot cavity. This increases the effective arm-length and thus
the sensitivity of the instruments. Gravitational waves (along with
seismic and other noise) change the relative length of the optical
cavities and produce an external strain,
\be
h(t)=\frac{L_x(t)-L_y(t)}{L_0}
\label{ht0}
\ee
that is incident on the interferometer. Here, $L_x(t)$ and $L_y(t)$
are the effective lengths of the $x$ and $y$-arms, respectively and
$L_0$ the effective length of the cavities in the absence of an
external strain.

The output of the interferometers, however, is not $h(t)$. Rather, the
output is derived from light that escapes the anti-symmetric
(``dark'') port and provides the gravitational wave read-out. It is
often called the error signal or ``gravitational wave channel''. Here
we will refer to it as $q(t)$.

The re-construction of the gravitational wave strain $h(t)$ from the
error signal $q(t)$ is an essential part of the data analysis. Since
the LIGO instruments are linear the gravitational wave strain incident
on the interferometer is a a linear functional of $q(t)$, namely,
\be
h(t)=\int R(t-t')q(t')dt',
\ee
where $R(t)$ is a suitable convolution kernel. Convolution in the time
domain is simple multiplication in the frequency domain so that
\be
h(f)=R(f)q(f),
\label{eq2}
\ee
where $x(f)$ denotes the Fourier transform of $x(t)$. The ratio of the
gravitational wave strain to the error signal is called the response
function $R(f)$. Finding the response function, or alternatively the
kernel $R(t)$, is the procedure referred to as the calibration.

So far LIGO has used a calibration implemented entirely in the
frequency domain \cite{ligocal}, i.e. using Eq.~(\ref{eq2}). On the
other hand, the British-German interferometric detector GEO600
\cite{GEO} has always produced a real-time calibrated $h(t)$ using
time-domain filters \cite{Martin}.  Some progress toward the
production of $h(t)$ for the LIGO interferometers has already taken
place \cite{soumya}. In this work we adapt the GEO time-domain
calibration method to the case of the LIGO interferometers.
           
In Section~\ref{simple} we provide a simple description of the length
sensing and control system of the LIGO interferometers, review the
frequency domain calibration procedure and formally show how the
strain can be reconstructed from the error signal of the
interferometer in the time domain. In Section~\ref{match} we describe
a method to track changes in the response using sinusoidal excitations
injected into the instrument. In Section~\ref{pipe} we explain how the
digital filters used to reproduce and invert the responses of various
parts of the length control system are created and show their
performance in the case of the Hanford 4km interferometer during the
second science run (S2).  We also describe our signal processing
pipeline. We conclude in Section~\ref{concl}.

\begin{figure}[htb]
     \centering
        \includegraphics[width=6 cm]{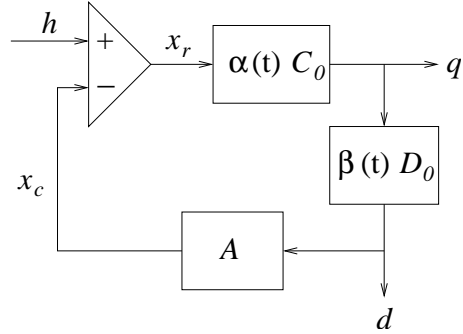}
        \caption{
Simple model for the feedback loop of the LIGO length sensing and
control system. The three filters that make up the feedback loop are
the sensing function $C=\alpha(t) C_0$, which converts the residual
strain $x_r$ into the error signal $q$, the feedback filter
$D=\beta(t) D_0$, which converts the error signal into the digital
control signal $d$, and the actuation $A$ which converts the digital
control signal into the control strain $x_c$ that is subtracted from
the external strain $h$. The digital signals $q(t)$ and $d(t)$ are
recorded by the data acquisition system at $16384\,{\rm Hz}$.
}
\label{simpleservo}
\end{figure}

\section{Simple description of the LIGO length sensing and control system}
\label{simple}

At low frequencies the external strain $h$ is dominated by seismic
noise and a control strain, $x_c$, is subtracted from it (see
Figure~\ref{simpleservo}) by physically moving the mirrors in the
Fabry-Perot cavity to compensate for the seismic noise.  This ensures
that the residual strain that enters the optical cavity, $x_r$, remains
small at low frequencies and keeps the optical cavities in resonance.

The optical cavity of the feedback loop is represented by the so-called
sensing function, $C=\alpha(t)C_0$. It converts the residual strain
$x_r$ into the digital error signal $q$, also called the gravitational
wave channel, which is sampled at $16384\,{\rm Hz}$ and recorded by
the data acquisition system. The error signal is measured in arbitrary
units called counts. Here, $C_0$ is a reference sensing function that
is measured at some time and $\alpha(t)$ is an overall (real) gain
that depends on the light power stored in the Fabry-Perot cavity which
changes with the alignment of the mirrors.  It is important to keep
track of this gain and the method we use is explained in
Section~\ref{match}. The frequency dependence of the sensing function
is determined primarily by the Fabry-Perot cavity in each arm and
corresponds to a real pole at around $90\,{\rm Hz}$.
 
A digital feedback filter, $D=\beta(t)D_0$, is applied to the error
signal $q$ which produces a digital control signal $d$. The digital
control signal, like the residual signal, is measured in units of
counts.  It is recorded by the data acquisition system at $16384\,{\rm
Hz}$. Here, $D_0$ is the feedback filter at some reference time and
$\beta(t)$ is a real overall gain.  The gain $\beta(t)$ is also
recorded by the data acquisition system and may vary in time. For the
first and second science runs (S1 and S2), however, it was kept fixed.
The filter $D_0$ consists of a double pole at DC and various low pass
filters as well as additional filters that keep the optics aligned and
the Fabry-Perot cavity in resonance.

The control signal $d$ is converted to strain and used to adjust the
length of the cavities via the actuation function $A$.  This is
achieved by converting the digital signal $d$ into currents that run
through coils placed near magnets glued on the back of the mirrors in
the Fabry-Perot cavity. The frequency response of the actuation
function is largely determined by the pendulum suspension of the
mirrors and thus consists of a pair of complex poles at the pendulum
frequency $f_p\approx0.75 {\rm Hz}$.

The re-construction of the external strain from the output of the
interferometer feedback loop is referred to as the calibration.
Calibration of LIGO data has thus far been performed in the frequency
domain by constructing a response function $R(f)$ that acts on the
error signal $q (f)$ of the feedback loop, namely,
\be
h(f)=R(f)\,q(f).
\label{hf}
\ee
The response function can be derived from the feedback loop equations
\begin{eqnarray}
\label{fback1}
x_r(f) &=& h(f)-x_c(f),
\\
\label{fback2}
q(f) &=& \alpha(t)C_0(f) x_r(f),
\\
\label{fback3}
x_c(f) &=& A(f)d(f),
\\
\label{fback4}
d(f) &=& \beta(t) D_0(f)q(f).
\end{eqnarray}
From these equations it is easy to show that 
$R(f)$ in Eq.~(\ref{hf}) is given by
\be
R(f)=\frac{1+\alpha(t)\beta(t)G_0(f)}{\alpha(t)C_0(f)},
\label{respf}
\ee
where $G_0=A C_0 D_0$ is the reference open loop gain. The optical
gain $\alpha(t)$ typically varies on a time-scale of seconds. The
digital gain $\beta(t)$ may also vary but was kept fixed during S1 and
S2. Hence Eqs.~(\ref{fback1}-\ref{fback4}) apply for frequencies above
a few tens of Hz.

We would like to perform this procedure entirely in the time
domain. To do this we reconstruct the strain time-series
$h(t)$ from the residual and control strains $x_r(t)$ and $x_c(t)$ as
follows. The residual strain $x_r$ is given by
\be
x_r(t)=h(t)-x_c(t),
\label{xr1}
\ee
so that
\be
\label{ht1}
h(t)=x_c(t)+x_r(t).
\ee
The low frequency part of the external strain is dominated by the
control strain $x_c(t)$ and the high frequency part is dominated by
the residual strain $x_r(t)$.

The optical gain of the sensing function $\alpha(t)$ typically varies
on much longer time-scales than the decay time of the impulse response
of the inverse of the sensing function. Thus, the residual strain can
be formally re-constructed according to
\be
x_r(t)\approx\frac{1}{\alpha(t)}T_{C_0^{-1}}[q(t)],
\label{xr2}
\ee
where $T_{C_0^{-1}}$ is a linear time-invariant filter operator for
the inverse of the sensing function $C_0$.

Similarly the control strain $x_c(t)$ can be constructed from $d(t)$
by filtering it through the actuation function
\be
x_c(t)=T_{A}[d(t)].
\label{xc1}
\ee
Therefore we can write the time series for the external strain in terms of
the error and digital control signals, $q(t)$ and $d(t)$ respectively,
as
\be
\label{ht2}
h(t)\approx\frac{1}{\alpha(t)}T_{C_0^{-1}}[q(t)]+T_{A}[d(t)].
\ee

Although $d(t)$ is an output of the interferometer servo loop recorded
by the data acquisition system it can be computed from the error signal
$q(t)$ again assuming that $\beta(t)$ varies in time much more slowly
than the decay time of the impulse response function of the feedback
filter\footnote{It should be noted that for the digital filter $D_0$
as it is currently defined this condition is never satisfied: The
double pole at DC results in an ill-behaved digital filter that does
not have a well-defined impulse response time. Later we show how to
construct a modified digital filter $D_0'$ which does satisfy this
condition.}. In particular,
\be
d(t)\approx \beta(t)\, T_{D_0}[q(t)].
\label{xc2}
\ee
So in terms of just the error signal $q(t)$ the external strain can
also be written as
\be
\label{ht3}
h(t)\approx\frac{1}{\alpha(t)}T_{C_0^{-1}}\left[ q(t) \right] 
	+T_{A}\left[ \beta(t)\,T_{D_0}\left[q(t)\right] \right] .
\ee
Note that, because $\beta(t)$ is constant we could have factored it
out of the second term on the right hand side of Eq.~(\ref{ht3}).

Thus, in order to produce the time-series for the external strain we
need 1) to track changes in the optical gain of the cavity, and 2)
construct time-domain digital filters for the inverse of the sensing
function, the actuation function and the feedback filter. These
problems are dealt with in the next two Sections.

\section{Tracking changes in the calibration}
\label{match}

In order to track changes in the optical gain of the cavity,
calibration lines are injected into the instrument by adding
sinusoidal excitations to the control signal.  This is a standard
technique used in gravitational wave detectors. In the LIGO
interferometers three calibration lines are injected; one at around
$1\,{\rm kHz}$, another one around $150\,{\rm Hz}$ near the unity gain
frequency of the open loop transfer function, and a third one at a few
tens of Hz.

\begin{figure}[htb]
     \centering
        \includegraphics[width=9 cm]{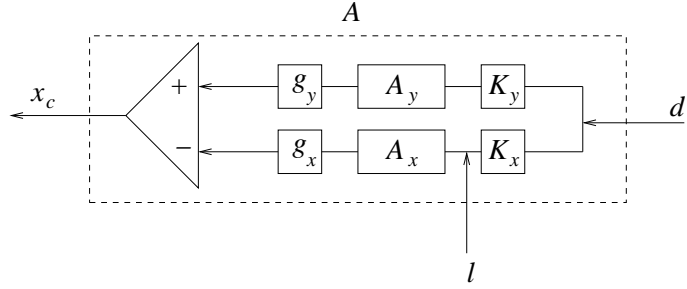}
        \caption{
Details of the actuation and calibration line injection point for the
LIGO interferometers length sensing and control system. This is the
content of the actuation function $A$ in Figure~\ref{simpleservo}.
}
        \label{actuation}
\end{figure}

The excitations may be added directly to the digital control signal
$d$. Alternatively they may injected into one of the two arms, as
shown in Figure~\ref{actuation}. During S2 they were injected into the
$x$-arm of the interferometers and for the purposes of the following
calculation we will assume this to be the case.  Generalisations to
injections into the $y$-arm or to the case where they are directly
added to the digital control signal $d$ are trivial.

Figure~\ref{actuation} depicts a more detailed model of the actuation
function $A$ that shows the injection point. The actuation function
$A$ is given by
\be
\label{actua}
A=  K_y g_y A_y- K_x g_x A_x,
\ee 
where $g_x$ and $g_y$ are analog gains measured at DC that depend on
the configuration of each of the arms and $K_x$ and $K_y$ are digital
gains that are mainly used to compensate for differences in the analog
gains. The functions $A_x$ and $A_y$ are the same for both arms except
for digital notch filters that take out frequency components of the
signal at the violin mode frequencies of the mirror suspensions. Since
the calibration lines are not injected at those frequencies, for our
purposes the functions $A_x$ and $A_y$ are identical along both arms
and we can write
\be
A_x=A_y \equiv A_*.
\label{actxyeq}
\ee

The excitation $l$ is added to the $x$-arm signal after the digital
gain $K_x$ is applied to the digital control signal. Therefore rather
than Eq.~(\ref{fback3}), the feedback equation for the control signal
at the frequency of a calibration line $f_c$ reads
\be
\label{xc_line1}
x_c(f_c)=A(f_c)d(f_c)-g_xA_*(f_c)l(f_c). 
\ee
Using Eqs.~(\ref{actua}) and (\ref{actxyeq}) we can write this as
\be
\label{xc_line2}
x_c(f_c)=A(f_c)[d(f_c)-\mu l(f_c)]. 
\ee
with $\mu=g_x/(K_y g_y - K_x g_x)$.  Since the calibration line is
injected with an amplitude that is large compared to the external
strain, the residual strain is dominated by the control signal at the
frequency of the calibration line. So to a good approximation we can
write the feedback equation for the control signal,
Eq.~(\ref{fback1}), as
\be
\label{xr_line1}
x_r(f_c)\approx -x_c(f_c).  
\ee
Combining Eqs.~(\ref{xc_line2}) and (\ref{xr_line1}) with the feedback
loop equations, (\ref{fback2}) and (\ref{fback4}) which remain
un-changed, we arrive at two independent expressions; one for the
product of $\alpha(t)$ and $\beta(t)$
\be
\label{alphabeta}
\alpha(t)\beta(t) \approx \frac{1}{G_0(f_c)}\frac{r_d(f_c)}{1-r_d(f_c)}
\ee
where $r_d=d/\mu l$, and another for $\alpha(t)$, 
\be
\label{alpha}
\alpha(t) \approx \frac{1}{A(f_c)C_0(f_c)} 
	\frac{r_q(f_c)}{1-r_d(f_c)}
\ee
where $r_q=q/\mu l$.

Thus, the sensing function gain $\alpha(t)$ as well as its product
with the feedback filter $\beta(t)$ are given by expressions involving
the complex functions $A$, $C_0$ and $G_0$ as well as complex ratios
of the error and digital control signals to the injected excitation.

The digital feedback filter gain $\beta(t)$ is recorded by the data
acquisition system and so Eqs.~(\ref{alphabeta}) and (\ref{alpha})
should be regarded as two independent ways of calculating the optical
gain $\alpha(t)$.

The ratios of the error and digital control signals to the injected
excitation are computed as follows. We take a fixed amount of time for
each of the three time-series $q(t)$, $d(t)$ and $l(t)$, and apply a
Hann window. We then perform a complex heterodyne at the frequency of
the calibration line $f_c$ and finally integrate over time. This
procedure yields complex $q(f_c)$, $d(f_c)$ and $l(f_c)$ which are
used to compute the ratios $r_d$ and $r_q$.

\begin{figure}[htb]
     \centering
        \includegraphics[width=12 cm]{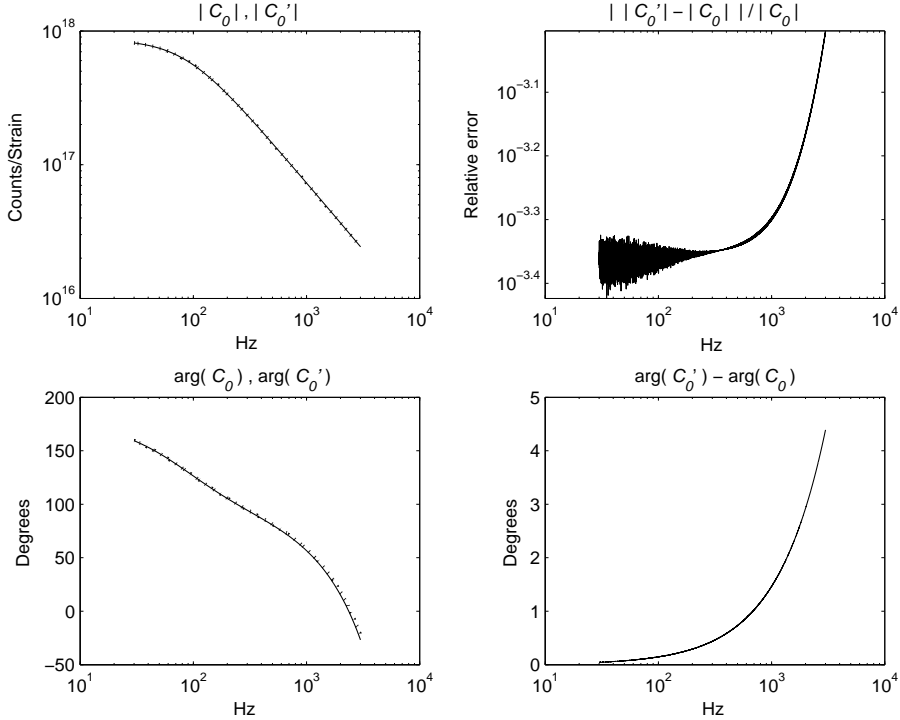}
        \caption{
The magnitude (top left) and phase response (bottom left) of the
analog sensing function $C_0(f)$ (solid curve) and the modified
digitised sensing function $C_0'(f)$ (dotted curve). On the top right
the absolute value of the relative error between the two magnitude
responses and on the bottom right the difference in degrees of the two
phase responses.
}
        \label{sense}
\end{figure}

Since the quantities in Eqs.~(\ref{alphabeta}) and (\ref{alpha}) are
complex and the measurement contains noise from the external
strain the optical gain $\alpha(t)$ has a small imaginary part.  This
imaginary part is small only to the degree to which
Eq.~(\ref{xr_line1}) is a good approximation. It should be noted
however that Eq.~(\ref{xr_line1}) can be made arbitrarily accurate by
either making the amplitude of the injected line arbitrarily large or,
since the external strain has no component that is correlated with the
calibration lines, by integrating for an arbitrarily long amount of
time.

The optical gain $\alpha(t)$ is typically calculated on time-scales of
a few tens of seconds which, given the typical injected amplitudes of
the calibration lines, is sufficient to reduce measurement noise
errors to an acceptable few percent.

\section{Implementation}
\label{pipe}

\subsection{Digitisation of filters and filter performance}

The three filters of the feedback loop can be described in terms of
analog zero and pole models. This is the starting point of the
digitisation procedure.

\begin{figure}[htb]
     \centering
        \includegraphics[width=12 cm]{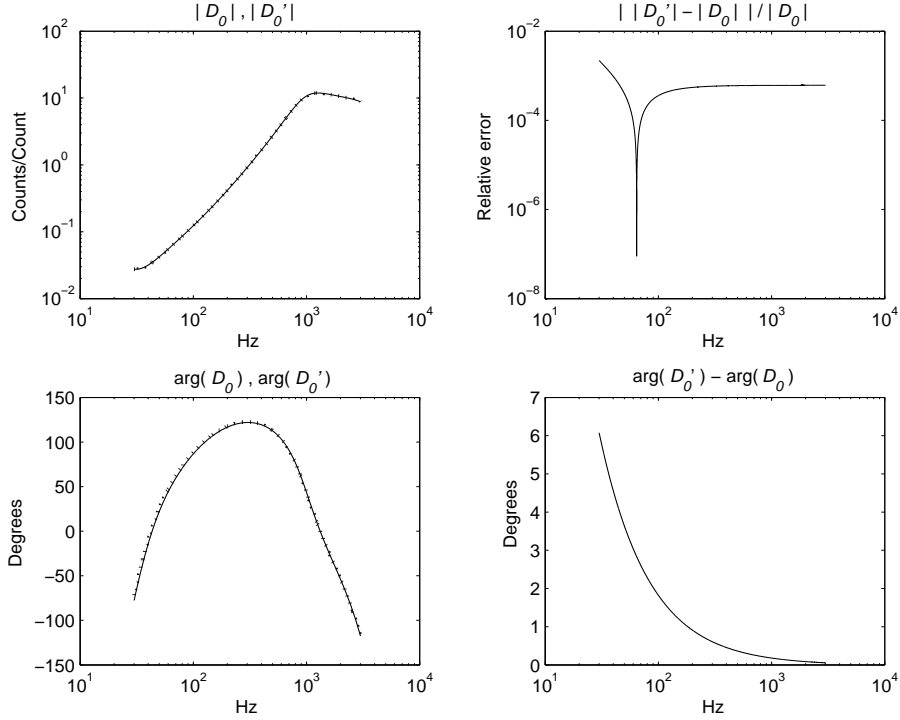}
        \caption{
The magnitude (top left) and phase response (bottom left) of the
original feedback filter $D_0$ (solid curve) and the modified feedback
filter  $D_0'$ (dotted curve). On the top right the absolute value of the
relative error between the two magnitude responses and on the bottom
right the difference in degrees of the two phase responses.
}
        \label{servo}
\end{figure}

The sensing function $C_0$, as we have already discussed, consists
essentially of a cavity pole at around $90\,{\rm Hz}$ so the inverse
of this filter is a zero at around $90\,{\rm Hz}$. Unfortunately,
digitisation of a zero does not lead to a well behaved digital filter.
In order to resolve this problem we introduce a pole at high
frequencies in the inverted sensing function $C_0^{-1}$ prior to
digitisation. To ensure the magnitude and phase response of the filter
do not change in the frequency range of interest\footnote{We chose
the range to be from $30\,{\rm Hz}$ to $3000\,{\rm Hz}$. This range
covers the frequency bands used in gravitational wave data analyses
performed by the LIGO Scientific Collaboration so far.} the additional
pole is placed at a sufficiently large value of the frequency. We then
digitise the analog filter using a bi-linear transform at a sample
rate consistent with the presence of the additional pole.

The sensing function also contains a unity gain $8^{\rm th}$ order
anti-aliasing elliptic filter at $7570{\rm Hz}$.  This filter
introduces frequency dependent phase shifts in the data that need to
be accounted for. The anti-aliasing filter contains a series of zeros
on the imaginary axis, which in the inverted filter turn into
poles.  To ensure that we obtain a well-behaved inverse digital
filter the zeros need to be moved away from the imaginary axis prior
to digitisation.

For the Hanford 4km interferometer in the S2 run we generated a
modified digital filter $C_0'$ with a bi-linear transform at a sample
rate $16$ times greater than the sample rate the of the time series
$q(t)$, i.e. $262144\,{\rm Hz}$. The zeros of the elliptic filter were
moved away from the imaginary axis by approximately $160\,{\rm Hz}$
and a single real pole was added at $10^5\,{\rm Hz}$.

\begin{figure}[htb]
     \centering
        \includegraphics[width=12 cm]{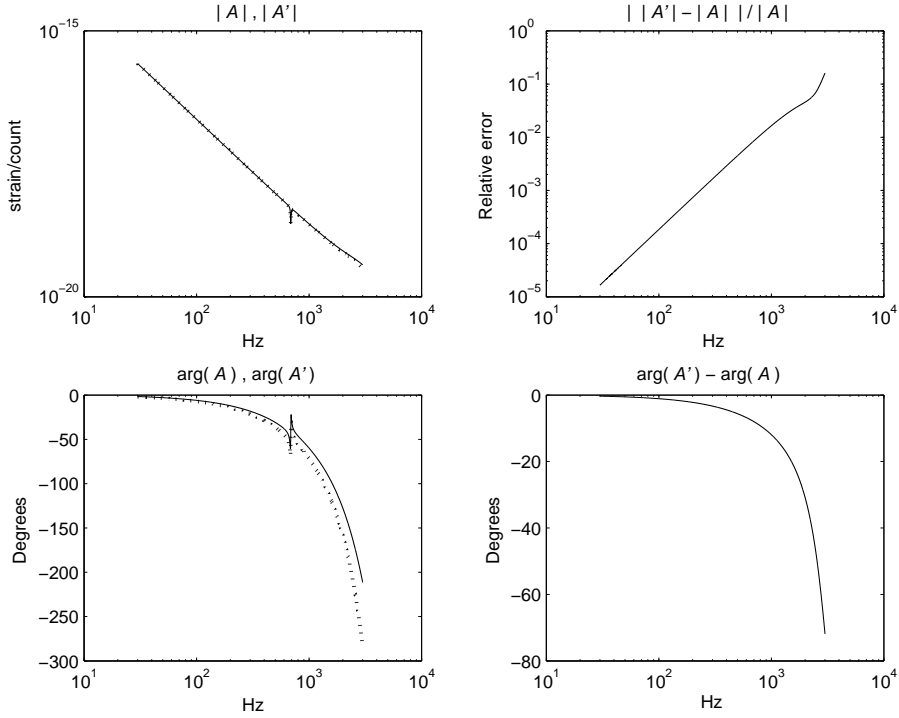}
        \caption{
The magnitude (top left) and phase response (bottom left) of the
original actuation function $A$ (solid curve) and the modified
actuation function $A'$ (dotted curve). On the top right the absolute
value of the relative error between the two magnitude responses and on
the bottom right the difference in degrees of the two phase
responses. The large errors in the phase beyond a few hundred Hz are
un-important because the contribution to the external strain $h(t)$
from the control signal $x_c(t)$ is negligible at these frequencies.
}
        \label{act}
\end{figure}

In Figure~\ref{sense} we show the magnitude and phase responses and
errors of the analog sensing function $C_0$, as well as our modified
digitised sensing function $C_0'$. The errors between $30{\rm Hz}$ and
$3{\rm kHz}$ are less than about $1\%$ in magnitude and $5^\circ$ in
phase.

The LIGO digital feedback filters $D_0$ are constructed from bi-linear
transformations at $16384\,{\rm Hz}$ of a series of analog filters.
The main one of these is a double pole at DC which, on its own outside
a feedback loop, is inherently unstable. In order to stabilise it, the
analog filter can be modified by shifting the double pole away from
zero by a small amount prior to digitisation. It should be noted that
this procedure changes the sign of the feedback filter at DC and
introduces errors in the response of the filter at low frequencies.
The rest of the filters in $D_0$ can be used exactly as they are
implemented in the instrument.

Figure~\ref{servo} shows the magnitude and phase responses, as well as
the errors, of the original digital feedback filter $D_0$ and the
modified digital feedback filter $D_0'$ for the Hanford 4km
interferometer during S2. As expected the errors are larger at low
frequencies. However, the errors between $30\,{\rm Hz}$ and $3\,{\rm
kHz}$ are less than about $1\%$ in magnitude and $6^\circ$ in phase.

The frequency response of the actuation function $A$ is primarily
determined by the pendulum suspension of the mirrors and thus consists
of a pair of complex poles at the pendulum frequency $f_p\approx 0.75
{\rm Hz}$. Additionally, the actuation function contains an
anti-imaging $4^{\rm th}$ order elliptic filter at $7570{\rm Hz}$, a
snubber used to suppress resonances in the electronics and a measured
$140\mu s$ time-delay modeled using a $4^{\rm th}$ order Pade
filter. The actuator also contains a series of digital notch filters
that remove components of the digital control signal at the violin
mode frequencies of the mirror suspensions.

All the analog parts of the actuator have been digitised using a
bi-linear transform at $16384\,{\rm Hz}$.  We have used the existing
digital notch filters as they are implemented in the instrument.

Figure~\ref{act} shows the result of this procedure. Plotted are the
magnitude and phase response of the original actuation function $A$
and the digitised actuation function used in the calibration procedure
$A'$ as well as the errors for the Hanford 4km interferometer during
S2. The errors in the magnitude and phase are less than about $1\%$
and $5^\circ$ respectively at frequencies up to a few hundred Hz. At
higher frequencies the errors in the phase become larger but this is
un-important: the contribution to the external strain $h(t)$ from the
control signal $x_c(t)$ is negligible above a few hundred Hz.

\subsection{Signal processing pipeline}

LIGO data is available in so-called ``science segments''. At these
times the cavities are in resonance and the instrument is sufficiently
stable to produce data of scientific quality.  The length of science
segments depends mostly on the level of seismic activity at the
location of each of the sites and can last anywhere between a few tens
of seconds to tens of thousands of seconds.

Science runs typically consist of a few hundred science segments. This
lends the calculation of the external strain for an entire science run
amenable to a cost-effective Beowulf cluster, which is an ensemble of
loosely coupled processors with a simple network architecture. In our
implementation each processor computes the external strain for a
science segment. The few hundred processes required can be scheduled
to run on the cluster with a batch system such as Condor
\cite{condor}. Indeed, the pipeline described below was used to
compute the external strain for the entire S2 run for all three LIGO
interferometers using Condor on the 300-node Medusa Cluster
\cite{medusa} at UWM.

\begin{figure}[htb]
     \centering
        \includegraphics[width=12 cm]{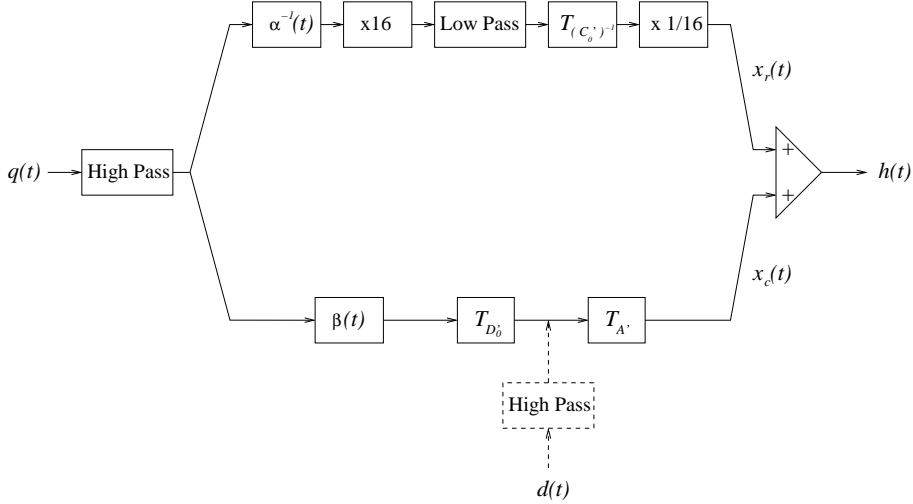}
        \caption{
Signal processing pipeline for time domain calibration.  Boxes labeled
$\times 16$ and $\times\,1/16$ denote up-sampling and down-sampling
respectively.
}
        \label{pipeline}
\end{figure}

The first step of our procedure is to compute a table of values for
the optical gain $\alpha(t)$ and the digital feedback filter gain
$\beta(t)$ for an entire science segment. As we have described in
Section \ref{match}, obtaining reliable values of the optical gain
requires an integration time of a few tens of seconds, so each science
segment will have anywhere from a few to a few thousand values of
$\alpha(t)$. For the Hanford 4km interferometer during S2 we chose to
compute the optical gain every 60s.

Science segments can be up to a few tens of thousands of seconds
long. Therefore we cannot keep all the data for a science segment in
the memory of a single node. As a result, in our signal processing
pipeline, each science segment must be divided into a number of
shorter parts. In the current implementation of the pipeline the data
is calibrated in $16$ second sub-segments.

Figure~\ref{pipeline} shows the digital processing pipeline used in
the computation of the external strain $h(t)$ from the error signal,
i.e. our implementation of Eq.~(\ref{ht3}).  The top and bottom
branches show the computations of the residual and control strains
respectively. 

To avoid problems involving the dynamic range of double precision
floating point numbers, we first high-pass filter 16s of the error
signal $q(t)$.  For the Hanford 4km interferometer during S2 we
implemented the high-pass filter using a $10^{\rm th}$ order
Butterworth filter at $40\,{\rm Hz}$. This is a conservative choice
and the specific frequency and order of the filter that is necessary
depends on the low frequency character of the data. Generally, we
expect these choices to depend on the specific science run and
interferometer.  For numerical stability the high-pass filter is
divided into second order sections. Furthermore, to ensure the phase
of the signal is not adversely affected by the filtering procedure,
the high-pass Butterworth filter is applied to each segment of the
data once forward and once in reverse. It should be noted that this
procedure is acausal but since we are not attempting to simulate a
physical process this is unimportant.

\begin{figure}[htb]
     \centering
        \includegraphics[width=12 cm]{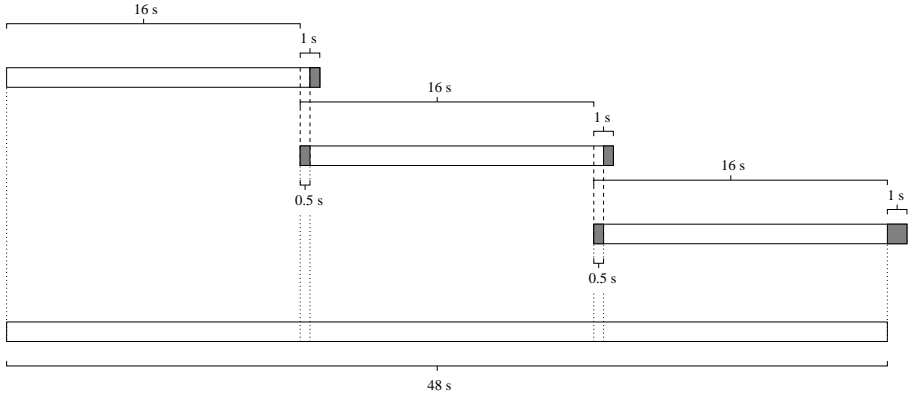}
        \caption{
Schematic of three consecutive 16s segments of data.  To avoid the
discontinuities produced by the finite impulse response of the
Butterworth filters in the pipeline, we filter an extra second of data
at the end of each 16~s segment. We then use the first half of this
second at start of the next segment. Shaded areas contain data for
which the filters have not settled and are replaced with data from the
previous or next segment.
}
        \label{segments}
\end{figure}

Since the data is being calibrated in segments care must be taken at
the boundaries between consecutive segments. The finite impulse
response time of the Butterworth filter produces a discontinuity at
the beginning and also at the end of each segment (because the
high-pass filter is applied forward as well as in reverse). To ensure
continuity across contiguous segments we must allow for extra time at
the end of each segment.  Figure \ref{segments} shows a schematic of
the procedure we have implemented in the case of three consecutive 16s
segments.  We use one extra second of data at the end of each segment
which allows sufficient time for the high-pass filter to settle. We
then then apply the first half of the extra second to the start of the
next segment of data. The shaded areas in Figure \ref{segments}
contain data for which the filters have not settled and are replaced
with data from the previous or next segment. It should be noted that
this procedure does not remove the few hundred milliseconds of initial
ringing of the Butterworth filter at the start of each science
segment.

To compute the residual signal $x_r(t)$ we first multiply each sample
of the high-pass filtered $q(t)$ by a suitably interpolated value of
the inverse of the optical gain $\alpha^{-1} (t)$. Interpolations are
necessary here because the filter gains are typically computed over a
time-scale of tens of seconds, and are not available for every sample
of the error signal $q(t)$. We use the table of values of the optical
gain created for the entire science segment described above to compute
an interpolated $\alpha^{-1} (t)$ for each sample of $q(t)$ using a
cubic spline. In principle we could also have used band-limited
interpolation but in practice this makes little difference.

The inverse sensing function filter has been digitised at
$262144\,{\rm Hz}$, and therefore the signal must be up-sampled by a
factor of $16$. The up-sampling procedure is simple: $15$ zeros are
added between samples and the resulting time-series is smoothed with a
low pass filter at a frequency lower than the Nyquist frequency of the
original time-series $q(t)$. For the Hanford 4km interferometer during
S2 we have chosen a $12^{\rm th}$ order Butterworth filter at $6\,{\rm
kHz}$ which is applied to each data segment once forward and once in
reverse in second order sections. To ensure continuity across
contiguous segments we apply a procedure along the lines described
above and in Figure \ref{segments} adding an extra second of data at
the end of each 16s segment.

We then filter the up-sampled signal through the inverse of the
modified sensing function $T_{(C_0')^{-1}}$.  We keep the history of
the digital filter across segments to ensure continuity. Finally, to
produce the residual signal $x_r(t)$ sampled at $16384\,{\rm Hz}$, we
down-sample by a factor of $16$ by picking out one out of every $16$
samples.

In our pipeline the digital control signal is computed by multiplying
each sample of the high-pass filtered time series $q(t)$ with an
interpolated value of the digital filter gain $\beta(t)$ and then
filtering the result by the modified digital feedback filter
$T_{D_0'}$. Up to differences between the original and modified
digital feedback filter this produces the digital control signal
$d(t)$. Note that we could have skipped these steps and instead used
the digital control signal that is read out from the instrument.

\begin{figure}[htb]
     \centering
        \includegraphics[width=12 cm]{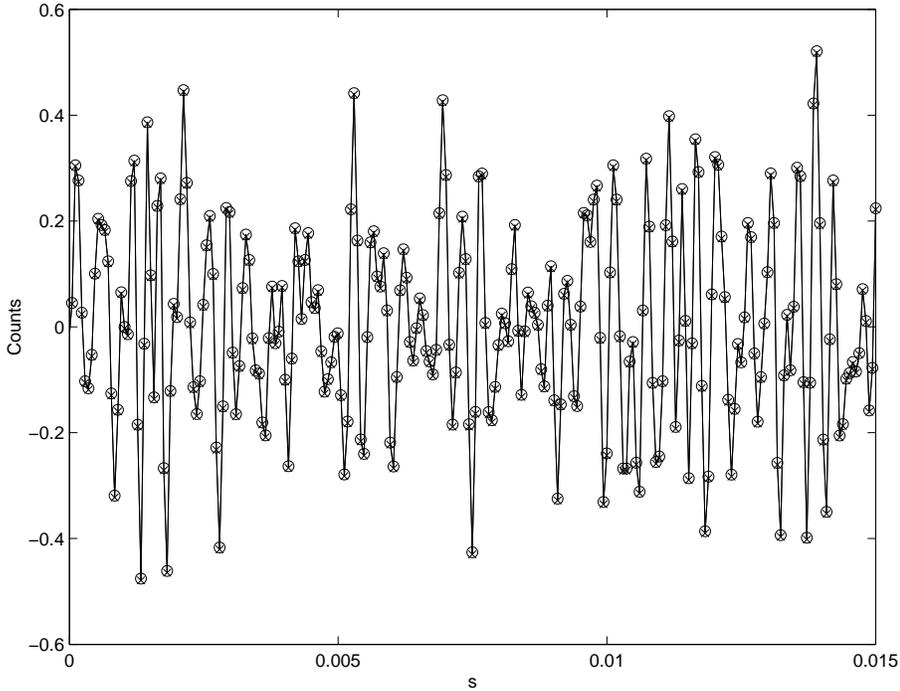}
        \caption{
Plot of the digital control signal $d(t)$ as recorded by the data
acquisition system (circles joined by a line) and the digital control
signal produced by our pipeline (crosses) versus time in seconds for a
segment of data taken by the Hanford 4 km interferometer during
S2. Due to the high-pass filtering done on the error signal $q(t)$
prior to filtering it through the modified feedback filter $T_{D_0'}$
the low frequency components of these two signals are different. To
allow for easier comparison both signals have been high-pass filtered
using $4$ second order Butterworth filters at $40\,{\rm Hz}$.
}
        \label{dplot}
\end{figure}

The low frequency components of $d(t)$ as read out from the instrument
and that computed by our pipeline are different. This is mostly due to
the high-pass filtering done on $q(t)$ prior to filtering it through
$T_{D_0'}$. In Figure \ref{dplot} we show a plot of the digital
control signal as recorded by the data acquisition system as well as
the one computed in our pipeline versus time for a segment of data
taken by the Hanford 4 km interferometer during S2. To allow for
easier comparison in Figure \ref{dplot} both the digital control
signal as recorded by the data acquisition system and the one produced
by our pipeline have been high-pass filtered by applying $4$ second
order Butterworth filters at $40\,{\rm Hz}$ consecutively. This
removes the low-frequency components and makes the agreement between
the two signals manifest.

The digital control signal is then filtered through the modified
actuation function $T_{A'}$ which produces the control strain
$x_c(t)$. The histories for both the digital feedback filter
$T_{D_0'}$ and the actuator $T_{A'}$ are stored in memory to ensure
continuity across 16s segments.

The calibrated strain $h(t)$ is then computed by summing the residual and
control strains $x_r(t)$ and $x_c(t)$.

\section{Conclusions}
\label{concl}

We have described a method to compute the strain $h(t)$ from the
output of the LIGO interferometers. This method has been implemented
off-line for all three interferometers for the S2 run and results
have been presented here for the Hanford 4 km interferometer.  The
codes that implement the pipeline as well as the digital filters are
available under the LSC Algorithm Library (LAL \cite{lal}) and the
procedure is fully automated under Condor \cite{condor}.

The pipeline introduces errors associated with the measurement noise
involved in the calculation of the optical gain $\alpha(t)$ as well as
the digitisation and modifications of the filters. We expect these
errors to be less than about $10\%$ in amplitude and $10^\circ$ in
phase across the detection band. We have tested the pipeline by
comparing Fourier transforms of time-domain calibrated data with data
calibrated in the frequency domain and found the differences to be
well within our errors. We have also faithfully reproduced the digital
control signal in the time domain from the error signal using our
modified servo filter $T_{G'_0}$ (see Figure \ref{dplot}).

At this time a third science run (S3) has taken place and a similar
off-line procedure will be applied to generate strain data for that
run. The current plan is to place an on-line system at the LIGO
sites to generate $h(t)$ in the near future.

\section*{Acknowledgments} 

We would like to thank Rana Adhikari, Gabriela Gonzalez and Daniel
Sigg for valuable discussions and suggestions.  X.~S.  would like to
thank the hospitality and support of the Albert Einstein Institute, at
which part of this work was done. The work of B.~A., J.~C. and
X.~S. was supported by National Science Foundation grants PHY 0071028,
PHY 0079683 and PHY 0200852.\\

\end{document}